\documentclass[doublecol]{epl2}

\usepackage{amssymb}
\usepackage{graphicx}
\usepackage{dcolumn}
\usepackage{bm}

\begin{document}

\title{Bands, resonances, edge singularities and excitons in core level spectroscopy investigated within the dynamical mean field theory}
\shorttitle{Bands, resonances, edge singularities and excitons in core level spectroscopy}

\author{M. W. Haverkort \inst{1,2,3} \and G. Sangiovanni \inst{4} \and P. Hansmann \inst{5} \and A. Toschi \inst{6} \and Y. Lu \inst{1,2,3} \and S. Macke \inst{1,2}}

\institute{
 \inst{1} Max Planck Institute for Solid State Research, Heisenbergstra{\ss}e 1, 70569 Stuttgart, Germany\\
 \inst{2} Department of Physics and Astronomy, University of British Columbia, Vancouver, British Columbia V6T1Z1, Canada\\
 \inst{3} Max Planck Institute for Chemical Physics of Solids, N{\"o}thnitzerstra{\ss}e 40, 01187 Dresden, Germany\\
 \inst{4} Institut f{\"u}r Theoretische Physik und Astrophysik, Universit{\"a}t W{\"u}rzburg, Am Hubland, D-97074 W{\"u}rzburg, Germany\\
 \inst{5} Centre de Physique Th{\'e}orique, CNRS, {\'E}cole Polytechnique, 91128 Palaiseau, France\\
 \inst{6} Institut f{\"u}r Festk{\"o}rperphysik, Technische Universit{\"a}t Wien, 1040 Vienna, Austria
}

\date{\today}

\abstract{
Using a recently developed impurity solver we exemplify how dynamical mean field theory captures band excitations, resonances, edge singularities and excitons in core level x-ray absorption (XAS) and core level photo electron spectroscopy (cPES) on metals, correlated metals and Mott insulators. Comparing XAS at different values of the core-valence interaction shows how the quasiparticle peak in the absence of core-valence interactions evolves into a resonance of similar shape, but different origin. Whereas XAS is rather insensitive to the metal insulator transition, cPES can be used, due to nonlocal screening, to measure the amount of local charge fluctuation.}



\maketitle

Core level photoemission (cPES) and core level x-ray absorption spectroscopy (XAS) have long been valuable tools in the field of material research for a huge range of compounds with a different degree of correlations \cite{deGroot:2008wo}. For example, XAS can probe the unoccupied density of states in GaAs, Al or Hydrocarbons \cite{Stohr:1982ug}, local properties of correlated $d$-shells in transition-metal compounds like the cuprates \cite{Thole:1985uc, Ghijsen:1988bm, Chen:1992ku}, or the atomic like ground state symmetry of rare earth ions in heavy Fermion compounds and impurities in an aluminum garnet used as laser medium \cite{Gunnarsson:1983uf, Thole:1985tz, Hansmann:2008kf, Willers:2011iy}. Interestingly, the same experiment seems to measure a different observable (empty density of states or local symmetry of the occupied wave-function) depending on the amount of correlations in the material. This dichotomy is also present in theory. The theoretical efforts for the description of cPES and XAS can roughly be divided into two approaches based on an itinerant or local starting point.

In the itinerant approach one approximates the interactions between electrons by a (mean-field) potential. As a result one obtains a set of freely moving particles. This is the case for Hartee-Fock or density functional theory (in the local density approximation) calculations. On this level of theory XAS is identified as the unoccupied single particle density of states and cPES as a delta-function representing the occupied core density of states\cite{Rehr:2000vn}. The core-valence interaction can be modeled as an additional potential that suddenly arrises after the photon absorption, leading to an edge singularity in the spectral function \cite{Roulet:1969co, Nozieres:1969tr, Doniach:1970wr, Ohtaka:1990vt}. For many systems, including most of the transition-metal and rare earth oxides, it has been realized early on that the inclusion of the explicit core-valence interactions beyond a mean-field potential is crucial. Many of these core level edges are much more determined by the local multiplet structure than by the band-structure of the material. Sawatzky and coworkers used local models approximating a solid by a single atom in an effective potential (crystal field) or by a small cluster (ligand field theory) \cite{Thole:1985tz,deGroot:2008wo}. These cluster models can be extended to include the band width of the material at the level of an Anderson impurity model\cite{Gunnarsson:1983vb, Zaanen:1985wk, vanderLaan:1986uwa, Fuggle:1988tt}.  For small clusters, the full quantum many-body problem can be solved with the use of exact diagonalization. In contrast to the independent particle approach the nature of the resulting spectra is intrinsically many-body and must not be confused with the unoccupied density of states.

 \begin{figure*}
    \includegraphics[width=1.0\textwidth]{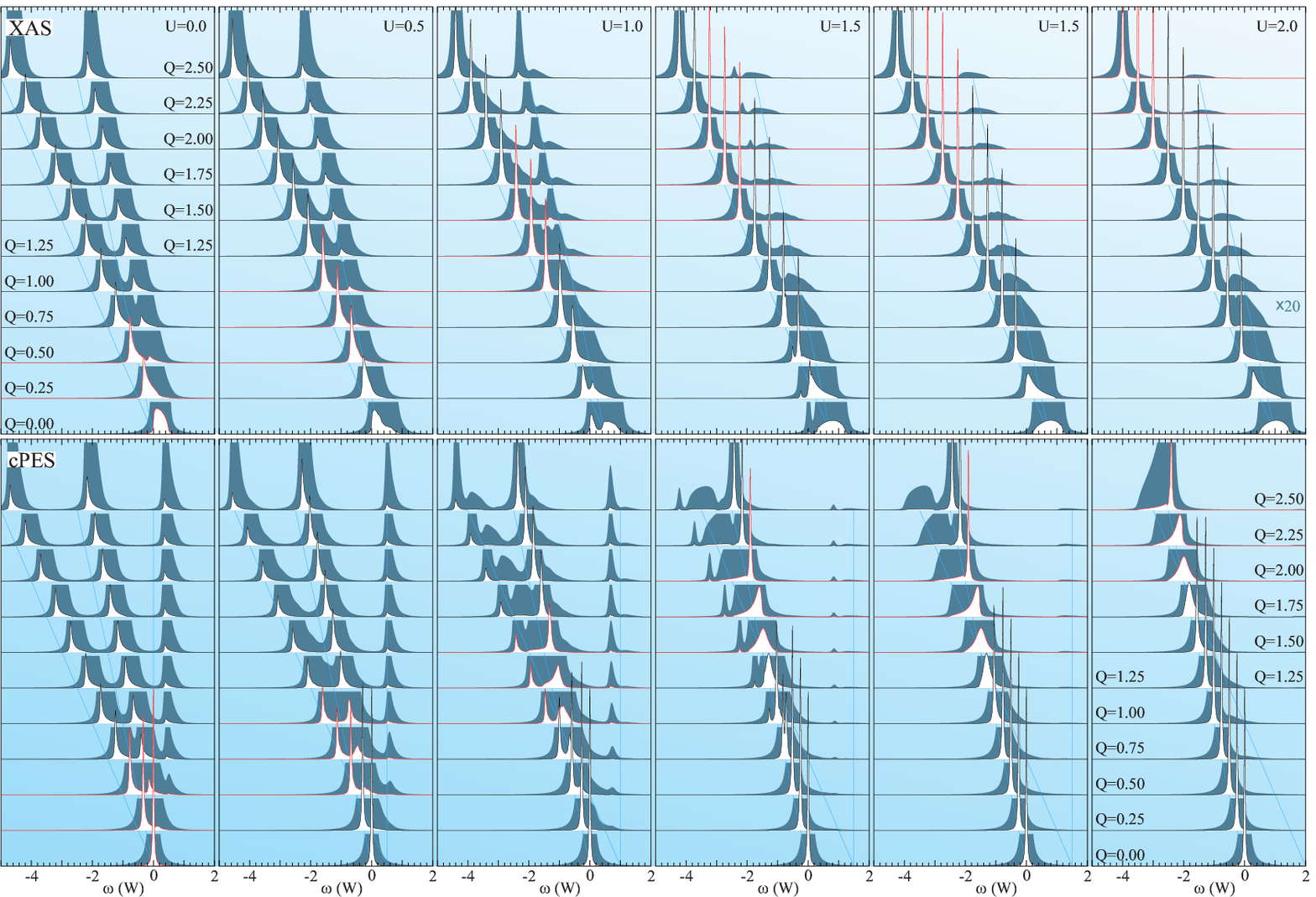}
    \caption{(color online) Top XAS and bottom cPES spectra calculated within the DMFT approximation for different values of the core-valence Coulomb interaction $Q=0$ (bottom) to $Q=2.5$ (top) and different values of the valence Coulomb interaction $U=0$ (left panel) to $U=2.0$ (right panel). There are two panels with $U=1.5$ either for a metallic ground state  (left) or an insulating ground state (right). Excitation energy $\omega$ is given in units of the one particle bandwidth $W$. The physical relevant curves where $U\sim Q$ are highlighted in red. The thin blue lines in the background show the excitation energies in the approximation where $W=0$. The shaded area shows the spectra multiplied by a factor of 20.}
    \label{Fig:All}
 \end{figure*}

Both viewpoints were initially quite disconnected, each including core level edges that could be successfully treated and edges which could not be described. Recently, there has been progress on both sides. For band insulators, one can start from density functional theory and use the Bethe-Salpeter equations for the calculation of XAS \cite{Laskowski:2010fv, Vinson:2011ep}. The local method developed into a full \textit{ab-initio} approach as a post Hartree-Fock or post density functional theory calculation similar to coupled cluster or configuration interaction schemes \cite{Ogasawara:2001kf, Ikeno:2005iq, Haverkort:2012du, Roemelt:2013ca}. Here we exemplify how dynamical mean-field theory (DMFT) \cite{Metzner:1989eq, Georges:1996un, Held:2007fi, Vollhardt:2012tl} can merge these two theoretical approaches for core level spectroscopy. 

DMFT is an approximation which locally includes all correlations and thus can reproduce the local models used for the interpretation of XAS and cPES \cite{Thole:1985tz,deGroot:2008wo}, including the full multiplet structure of the core-valence interactions. At the same time DMFT includes the full nonlocal band-structure on a  mean-field (Hartree-Fock or density functional theory) level as needed for the interpretation of XAS of less correlated materials. DMFT seems currently the perfect method for core level spectroscopy \cite{Cornaglia:2007kn, Sipr:2011ih, Hariki:2013hf}, merging the two separate theoretical approaches for XAS and cPES that exists to date. DMFT is computationally nontrivial as it requires to solve an Anderson impurity model. The solution to an Anderson impurity model which includes several correlated orbitals per site as well as spin-orbit coupling, low symmetry crystal-fields and the full rotational invariant Coulomb interaction is a complicated task. Here we use a recently introduced impurity solver, which allows one to do all this with the use of exact diagonalization \cite{Lu:2014de}. The bath or band-structure is approximated by 301 poles, leading to a many-body problem of about $10^{179}$ single Slater determinant basis functions. By a rotation to the natural orbitals of the impurity problem the number of important determinants that need to be stored can be reduced to a tractable number and the calculation of core level spectroscopy becomes possible with only moderate computational costs \cite{Haverkort:2012du, Lu:2014de}. 

In this letter, we exemplify the effect of core-valence interactions by calculating XAS and cPES spectra for different core-valence as well as valence-valence Coulomb interactions at zero temperature. We show the spectral function for a metallic, correlated metallic and Mott insulating ground state. As a model we take a single band system with a semi-circular density of states and onsite Coulomb interaction. One can think of these calculations as the $L_{3}$ edge of a Cuprate.\footnote{For real materials the interpretation of experimental XAS and cPES spectra are generally not captured by a single band Hubbard model, due to other bands that play a role in describing the electronic structure such as the O-$2p$ derived bands in transition metal compounds.} The ground-state has on average a single hole per Cu site in the $d_{x^2-y^2}$ band. The excited state for cPES has an additional core hole in the Cu-$2p$ shell. The excited state for XAS differs from the excited state of cPES by the fact that there is one additional electron in the Cu-$3d_{x^2-y^2}$ band. Using a mean-field Hartree-Fock or density functional approximation, where electrons behave as independent particles, these spectra are described by a single electron excitation from the Cu-$2p$ shell (orbital) to the Cu-$3d$ band (XAS) or to one of the free electron like states (cPES). In a local many-body language (neglecting charge fluctuation in the notation) XAS (cPES) is given as an excitation from a $|2p^6\,3d^9\rangle$ ground state to a $|2p^5\,3d^{10}\rangle$ ($|2p^5\,3d^9\rangle$) excited state.

In Fig. \ref{Fig:All} we show the XAS (top) and cPES (bottom) calculations. The different panels show calculations for different values of the valence Coulomb interaction ($U$). The Coulomb repulsion is changed from $U=0$ in the left panel to $U=2.0$ in the right panel. The different spectra within a panel show calculations for different core-valence Coulomb interaction ($Q$). The core-valence Coulomb interaction is varied from $Q=0$ for the bottom spectra to $Q=2.5$ for the top spectra. All interactions are given in units of the band-width. Our results obtained with our newly implemented solver reproduce well the cPES spectra calculated by Cornaglia and Georges \cite{Cornaglia:2007kn}.

At $Q=0$ (bottom spectra of all panels) the core and valence states decouple such that cPES measures a delta function and XAS is up to differences due to matrix elements involved, equivalent to inverse photo electron spectroscopy (IPES). The evolution of the spectra with increasing $U$ shows the well known transfer of spectral weight from the coherent quasiparticle peak centered on $\omega=0$ to Hubbard bands at $\pm U/2$ (in IPES one only observes the part for $\omega>0$) \cite{Metzner:1989eq, Georges:1996un, Held:2007fi, Vollhardt:2012tl}. For $U \lesssim 1.2$ DMFT finds a metallic and for $U \gtrsim 1.5$ a Mott insulating ground state. For parameters in the coexistence region ($1.2 \lesssim U \lesssim 1.5$) DMFT finds both an insulating and metallic solution.\footnote{The exact values of the coexistence region depend within a few percent on the algorithm used \cite{Blumer:2003tz}.} We show for $U=1.5$ in Fig. \ref{Fig:All} both cases. For $Q=0$ (and only for $Q=0$) a Fermi energy at $\omega=0$ in the sense of a photoemission experiment can be identified. For finite $Q$ there is no unique well defined way to relate the XAS spectra to the empty density of states. 

 \begin{figure}
    \includegraphics[width=0.49\textwidth]{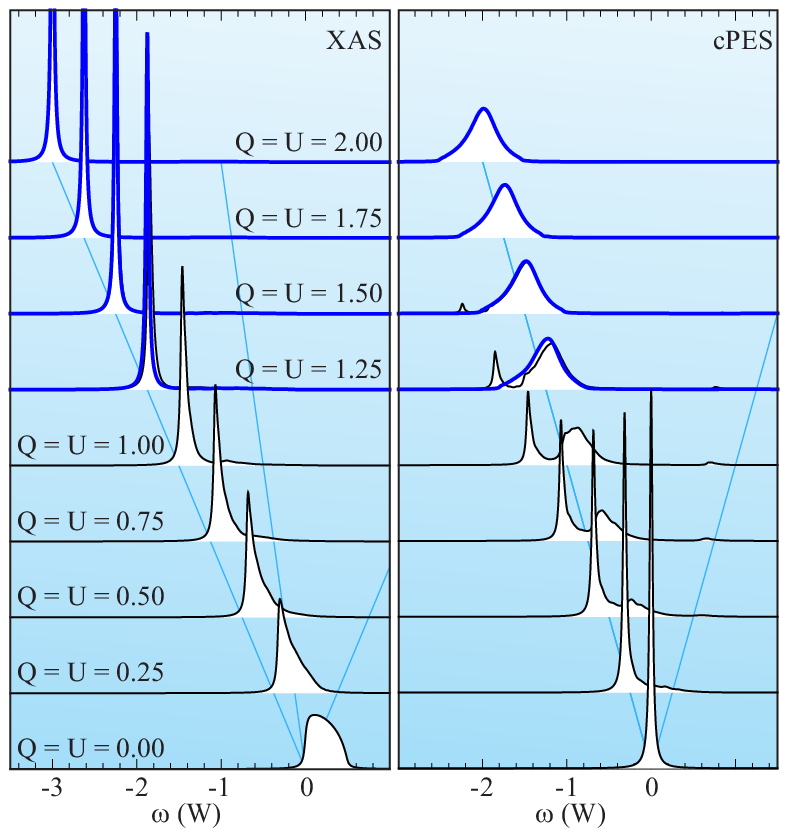}
    \caption{(color online) XAS (left panel) and cPES (right panel) calculated for different values of $U=Q$. Spectra for a metallic ground state are shown with thin (black) lines, spectra for an insulating ground state with thick (blue) lines.}
    \label{Fig:UisQ}
 \end{figure}

If one introduces core-valence interactions the XAS spectra are no longer equivalent to the IPES. For all $U$ values, by increasing $Q$ one observes a shift of the spectral intensity towards lower energy which sharpens at the onset, (edge singularity \cite{Roulet:1969co, Nozieres:1969tr, Doniach:1970wr, Ohtaka:1990vt}). For an insulating ground state and large enough $Q$ the lowest excitation is excitonic, whereas for metals the first peak is always a resonance. For nonzero $U$ the cPES is no longer given by a single delta function. For $U=0$ (left column) and $Q=2.5$ (top spectrum of each panel) one can see two (three) separated structures in the XAS (cPES) spectra. If one increases $U$ ($U<Q$) the three (two) peak structure in cPES (XAS) remains, but the spectral weight is shifted.

The XAS for nonzero $Q$ is not directly related to the unoccupied density of states or IPES. For correlated metals the coherent quasiparticle peak found in IPES is destroyed in XAS and replaced by an edge resonance due to the core-valence interaction. Nonetheless these spectra contain a wealth of information about correlations and local charge fluctuation. In order to understand how one can obtain this information from these spectra we discuss the full initial and final state wave-function from a local perspective. The many-body ground-state wavefunction ($\psi_0$) can be written as:
\begin{equation}
\psi_0=\alpha \psi_{d_{x^2-y^2}^0 , r^{N+1} } + \beta  \psi_{d_{x^2-y^2}^1 , r^{N} } + \gamma  \psi_{d_{x^2-y^2}^2 , r^{N-1} },
\end{equation}
with each term having $0$, $1$ or $2$ electrons at the site where the core-hole will be created and $N+1$, $N$ or $N-1$ electrons in the rest ($r$) of the solid. The amount of correlations is related to the local charge fluctuation, which are uniquely defined by $\alpha$, $\beta$, and $\gamma$. The average occupation ($\langle n \rangle$) is $\beta^2+2\gamma^2=1$, the double occupation is $\gamma^2$ and the charge fluctuation defined by the variance of the local occupation ($\sigma^2=\langle n^2\rangle -  \langle n\rangle^2$) is $2\gamma^2$. For $U=0$ we have $\alpha^2=\gamma^2=1/4$, $\beta^2=1/2$, and $\sigma^2=1/2$. For nonzero $U$ the charge fluctuation is reduced, and $\beta^2$ tends to one. It is this local occupation and its fluctuation, that one can probe with core level spectroscopy. In order to understand how, one can look at the many-body final state after the core hole is created. The different final states ($\psi_i$) can be written as:
\begin{equation}
\psi_i=\alpha_i \psi_{\underline{c}d_{x^2-y^2}^0 , r^{N+2} } + \beta_i  \psi_{\underline{c}d_{x^2-y^2}^1 , r^{N+1} } + \gamma_i  \psi_{\underline{c}d_{x^2-y^2}^2 , r^{N} }.
\end{equation}

The relation between the charge fluctuation in the many-body ground state and the cPES (XAS) spectra is easy to understand in the limit where $Q \gg W$ and $Q > U$. In this limit the many-body final states consist of three independent sets with either 0 ($\alpha_i=1$), 1 ($\beta_i=1$) or 2 ($\gamma_i=1$) electrons on the site of the core-hole. The spectral weight of these peaks is by a simple sumrule equal to the squared occupation numbers $\alpha^2$, $\beta^2$ and $\gamma^2$ of the ground-state, the energy of these peaks is related to $U$ and $Q$. In cPES the $d_{x^2-y^2}^0$ part of the wave-function is excited to the set of peaks at $U$, with weight $\alpha^2$, the $d_{x^2-y^2}^1$ part of the wave-function is excited to the set of peaks at $-Q$, with weight $\beta^2$, and the $d_{x^2-y^2}^2$ part of the wave-function is excited to the set of peaks at $U-2Q$,  with weight $\gamma^2$. For XAS the $d_{x^2-y^2}^2$ part of the wave-function cannot be excited so one only obtains two set of peaks instead of 3. The energies of the XAS peaks are $U-Q+\epsilon_d$, and $U-2Q+\epsilon_d$. These energies are indicated by the light blue lines in Fig. \ref{Fig:All}. 

Considering the previous discussion it might be surprising that for $U=0$ and $Q=2.5$ the intensity ratio is not exactly 1:2:1. This is related to the fact that $Q=2.5$ is not infinitely larger than $W=1$. The final states are not given by $\alpha_i=1$ ($\beta_i=1$, $\gamma_i=1$) for the first (second, third) excited state but by a mixture of these three basis states. In this case the phase relations between the different terms of the wavefunction play a crucial role in determining the spectral intensity. Generally it turns out that states at a lower energy gain spectral weight compared to the $Q\to\infty$ limit, as is also observed in Fig. \ref{Fig:All}. A clear example and an analyzes of this phenomena is given by a model calculation of a Li$_2$ molecule \cite{Sawatzky:1980ik}.

 \begin{figure}
    \includegraphics[width=0.49\textwidth]{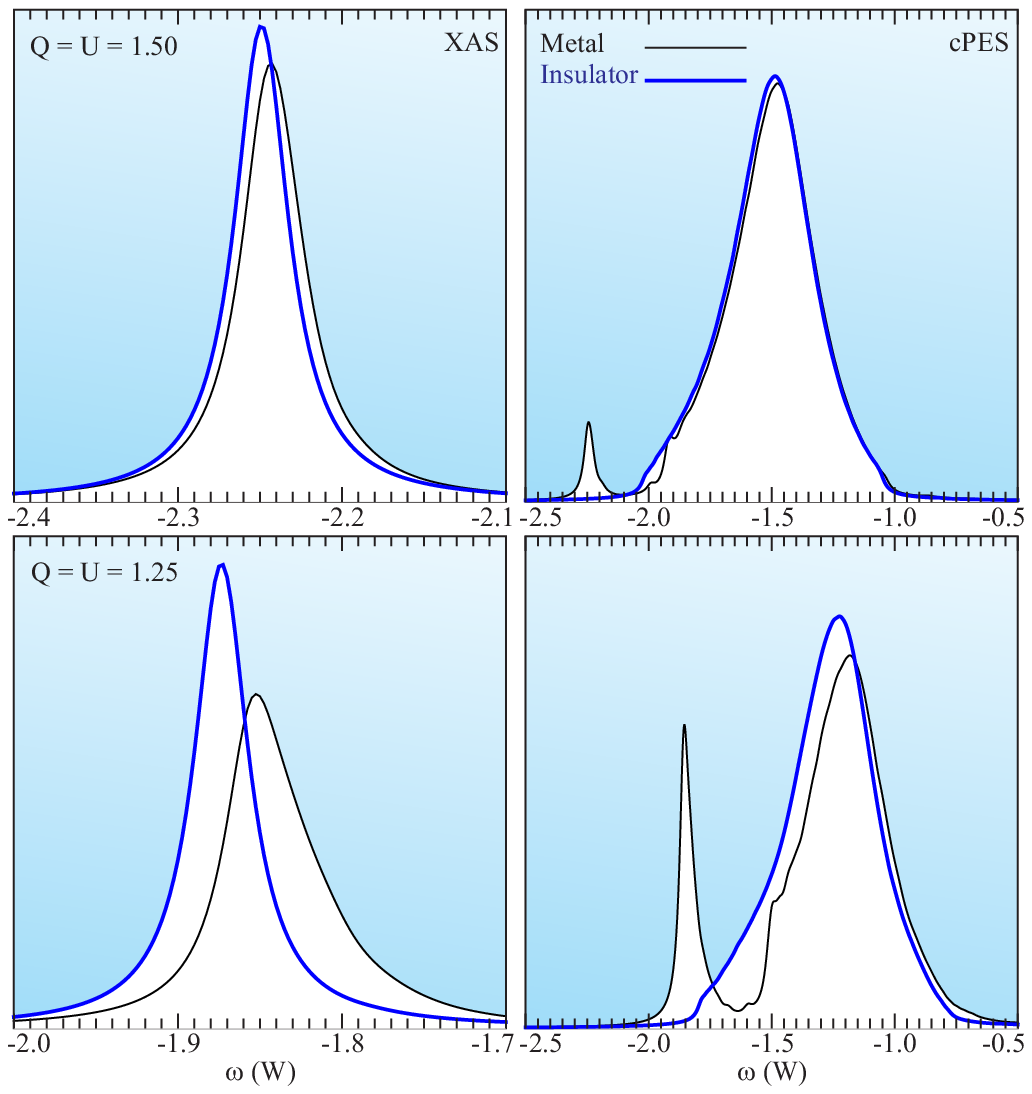}
    \caption{(color online) XAS (left panels) and cPES (right panels) for $U=Q=1.5$ (top) and $U=Q=1.25$ (bottom) comparing the spectra for a metallic (thin black) and insulating (thick blue) ground state. }
    \label{Fig:UisQMIT}
 \end{figure}

One cannot tune $Q$ in the experiment and some of the spectra in Fig. \ref{Fig:All} hardly present a realistic situation for the measurement of a material. At the transition metal $L_{2,3}$ edges, where one excites a $2p$ core electron one generally finds $Q\sim 1.2 U$ \cite{Bocquet:1996vt}. A closer look to these physically relevant spectra can be seen in Fig. \ref{Fig:UisQ} where we show the spectral evolution as a function of $U$ with $Q=U$. 

For $U=Q$ the XAS looks particularly simple and is almost given by a single peak, which is asymmetric (resonance) for a metallic ground state and a single peak (exciton) for an insulating ground state. For increasing $U$ the edge resonance becomes more and more symmetric and there is almost no change at the phase transition between the metallic (black) and insulating calculations (blue) in Fig. \ref{Fig:UisQ}. The cPES for $U=Q$ shows more features than the XAS. In the limit where $Q \gg U$ one would expect three peaks, one at $U$ (locally $d_{x^-y^2}^0$), one at $-Q$ (locally $d_{x^-y^2}^1$), and one at $U-2Q$ (locally $d_{x^-y^2}^2$). When $U=Q$ the last two peaks are degenerate and when hopping is included form, with the rest of the solid, a band. This is the broad feature seen in the right panel of Fig. \ref{Fig:UisQ} for large $Q=U$. 

Besides this band like feature in cPES an additional sharp peak appears for a metallic ground state. This is the edge singularity and is related to nonlocal screening \cite{vanVeenendaal:1993ve, vanVeenendaal:1993vq, Ide:2000ey}. The lowest cPES state is one where the core hole is screened by an additional electron and the aditional hole moves freely in the solid. For larger $U$ the intensity of this peak decreases and becomes less asymmetric. (The critical exponent of the edge singularity changes). For insulating solutions the edge singularity disappears as charge excitations are gaped. One should note that this behavior is not well reproduced if one calculates the spectra from a model consisting of only two sites (Li$_2$ molecule). 

In DMFT the (T=0) metal to insulator transition in the Hubbard model is a first order phase transition. DMFT can find a metallic or insulating self-consistent solution for some range of $U$ ($1.2 \lesssim U \lesssim 1.5$) (see also footnote 2). In Fig.~\ref{Fig:UisQMIT} we consider the XAS and cPES spectra for this regime. We show in thick blue the spectra calculated for an insulating ground state and in thin black the spectra calculated for a metallic ground state. The parameters are the same, for both calculations, the difference is solely in the ground state. The cPES spectra are clearly sensitive to these changes, whereas the XAS spectra show hardly any difference. Experimentally this can be seen for example at the metal insulator transition (MIT) of VO$_2$ \cite{Eguchi:2008fq, Haverkort:2005cr}, V$_2$O$_3$ \cite{Park:2000vv, Rodolakis:2010il, Hansmann:2012it} or Ca$_2$RuO$_4$ \cite{Kim:2004gx}. The polarization dependent XAS is sensitive to the local orbital occupation, which changes at the MIT for these materials. The isotropic XAS looks rather similar for the metallic and insulating phase \cite{Park:2000vv, Haverkort:2005cr, Rodolakis:2010il, Hansmann:2012it}. The cPES on the other hand is sensitive to the amount of charge fluctuation of the ground state due to the existence of a sharp edge resonance which clearly stands apart from the rest of the spectra. The intensity of this first peak is directly related to the amount of charge fluctuation in the ground state (see also footnote 1).  

In conclusion we show, using a recently implemented exact diagonalization impurity solver \cite{Haverkort:2012du,Lu:2014de}, how DMFT captures excitons, resonances, edge singularities and band excitations in core level spectroscopy. As DMFT can capture both the full local many-body interactions, including the core-valence interactions as well as the non-local bandstructure, it merges two different theoretical methods presently available for the calculation of XAS and cPES. For near edge features this method should describe the core level spectra rather well. Excitations above the continuum edge will require large basis sets, including many of the unoccupied bands and are probably described with less computational effort using multiple scattering techniques \cite{Rehr:2000vn}. 

It will be interesting to apply DMFT, using basis sets including both the correlated $d$ bands as well as the ligand (O $2p$) bands, on realistic materials \cite{Sipr:2011ih, Hansmann:2012it, Hariki:2013hf}. In that case core level spectroscopy can be used to critically test \textit{ab initio} DMFT calculations. Thereby not only testing the experimental realization of the low energy cluster parameters obtained from the \textit{ab initio} calculation, but also the DMFT approximation of treating non-local correlations on a mean-field level. Using cPES one could, for example, compare the amount of charge fluctuations obtained in \textit{ab initio} DMFT to experimentally observed values.

We would like to thank George Sawatzky and Antoine Georges for stimulating discussions. Financial support by the Deutsche Forschungsgemeinschaft through FOR 1346 (FWF project I-597-N16, AT) is gratefully acknowledged.

\bibliographystyle{eplbib}

\end{document}